# Sheath Analysis of Conducting Channel Walls in Anode-Layer Hall Thruster with Acceleration Zone Shifted Downstream

Rei KAWASHIMA*

*Department of Electrical Engineering, Shibaura Institute of Technology

**Abstract**

*Channel wall sheath thickness in an anode-layer Hall thruster RAIJIN66 was investigated by a two-dimensional hybrid simulation and a one-dimensional sheath model. A hybrid particle-fluid model with the quasineutrality assumption was used to obtain the wall ion flux and plasma property at plasma−sheath edge. The plasma property inside the sheath was estimated by the nonneutral sheath model. The simulated ion wall current density at the channel exit was 100 A/m$^2$, and the estimated sheath thickness was 0.64 mm. If the wall ion current density is reduced to 1/20 of this value, the sheath region will be extended to cover 50% of the channel width.*

## 1. Introduction

Hall thruster is one of the electric propulsion systems for spacecrafts. This thruster has an annular discharge channel where a crossed-field configuration of radial magnetic and axial electric fields is applied. Anode-layer type Hall thruster (thruster with anode layer: TAL) is one type of Hall thruster, and it exhibits different features from the magnetic-layer type Hall thruster (stationary plasma thruster: SPT) [1]. Due to its short discharge channel, the TAL is characterized by dense plasma generation and small thruster body. The TAL has conducting channel walls biased at the cathode potential to reduce the electron losses to the channel walls. One of the challenges in developing a TAL is that a TAL typically suffers from severe discharge current oscillation compared with SPTs [2]. Although the TAL has potential for efficient plasma generation, its superiority compared with the SPT has not been demonstrated [3].

In Japan, a research project to develop high-performance TAL thruster named "RAIJIN" is in progress. The 4.5-kW class RAIJIN94 attained the thrust efficiency of 0.60, which was comparable with the efficiencies of SPTs of similar input power levels [4]. A 2-kW RAIJIN66 was also designed where the discharge channel and magnetic field geometries were similarly scaled from those of RAIJIN94 so that the discharge plasma properties should be similar [5]. These thrusters are characterized by magnetic fields where the peak of magnetic flux density is located downstream of the channel exit. Owing to this magnetic field design, the RAIJIN thrusters exhibited primary acceleration zones that shifted downstream beyond the channel exit.

In a TAL with the shifted acceleration zone, the sheath structure in front of the channel walls is concerned in terms of energy efficiency and thruster lifetime. The typical potential structures in an SPT, TAL, and TAL with shifted acceleration zone are shown in **Fig. 1**. The discharge plasma in a conventional TAL forms a steep potential gradient in front of the anode, which is the origin of the name "anode layer" [6]. In the case of TAL with shifted acceleration zone, the plasma potential inside the discharge channel is maintained near the anode potential. Because the channel wall is biased at the cathode potential, it is presumed that a thick sheath is formed with a potential drop. This sheath attracts ions to the conducting channel walls, leading to wall ion loss. The wall ion loss means a loss of energy for ionization and acceleration. Further, the ions colliding with the walls have high energy owing to the large potential drop of the sheath, which causes wall erosion due to ion bombardment. Wall erosion is a major concern in Hall thruster systems because it is known as a limiting factor of thruster lifetime. Therefore, the sheath structure is closely related to the Hall thruster performance.

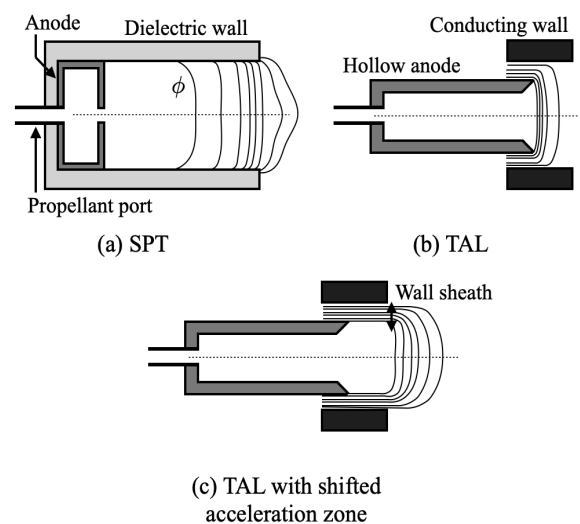

**Fig. 1** Presumed equipotential lines in typical operations of (a) SPT, (b) TAL, and (c) TAL with shifted acceleration zone.

In Ref. 5, the plasma properties inside RAIJIN66 TAL were measured, and it was observed that the sheath and presheath do not affect the bulk discharge plasma around the channel centerline. However, the sheath thickness was not experimentally evaluated, because plasma diagnostics of sheath regions close to the walls are technically challenging due to the small sheath thickness that is typically less than 1 mm.

In this study, a numerical model analysis is conducted as an alternative approach to revealing the sheath thickness in the TAL with shifted acceleration zone. A particle-fluid hybrid model is used for the bulk plasma flow, and a one-dimensional (1D) sheath model is developed to estimate the sheath thickness of the conducting channel walls.

## 2. Hybrid PIC and Electrostatic Sheath Models
### 2.1 Hybrid model for bulk discharge plasma

Several numerical simulations have been conducted for the discharge in TALs. Because the discharge in conventional TALs exhibits the nonneutral anode layer where the ion acceleration takes place, non-neutral full particle-in-cell (PIC) models have been usually employed for discharge modeling. Szabo conducted a full PIC simulation for a small cylindrical TAL where the channel diameter was 4.8 mm [7]. Yokota et al. simulated an annular TAL, to investigate the role of hollow anode for discharge stabilization [8]. Tang et al. and Geng et al. used a commercial full PIC software to simulate a three-dimensional flow in a cylindrical TAL [9,10]. The number of numerical works using full fluid or hybrid particle-fluid models is limited. Keidar and Boyd used a fully fluid model with a quasineutral assumption for simulating a high-power TAL; however, this model was focused on a downstream region, and the main ionization and acceleration zones in the anode layer were not simulated [11].

In the TALs with shifted acceleration zones downstream, the acceleration zone exists in the plume region and its length in the axial direction is much larger than the Debye length. In this sense, one can assume that an anode layer no longer exists, and that the discharge plasma is quasineutral except for the channel wall sheaths. For simulating the quasineutral plasmas such as the ones in SPTs, the hybrid particle-fluid model with the quasineutral assumption has been popularly used [12].

In this study, the hybrid PIC method is used where the ions and neutral atoms are treated as particles while the electrons are assumed as fluid. The model basics for the axial-radial domain of the Hall thruster has been explained elsewhere [13]. The quasineutrality is assumed in this model, and hence the non-neutral regions such as sheath are not included in the calculation domain. The ion and neutral particle motions are described by the equation of motion, which is computed by using a second-order time accurate leap-frog method. The electron fluid model consists of the conservations of mass, momenta, and energy in the two dimensions. The electron fluid equations are calculated by applying second-order space accurate schemes [14].

### 2.2 Cross-field electron transport model

Many models have been proposed for the cross-field electron transport in the orthogonal direction of magnetic lines of force. The most standard model is to include an anomalous electron transport effect in addition to the mobility obtained from the classical diffusion theory. It is considered that the anomalous electron transport is induced around the channel exit and plume region [15]. The anomalous electron transport is often reflected in numerical simulations by artificially increasing the scattering collision frequency of electrons. The anomalous collision frequency is modeled based on the Bohm diffusion as [16]

$$\nu_{\text{ano}} = \frac{\alpha_B}{16} \cdot \frac{eB}{m_e}, \qquad (1)$$

where $\nu_{\text{ano}}$, $e$, $B$, $m_e$ are the anomalous collision frequency, elementary charge, magnetic flux density, and electron mass, respectively. $\alpha_B$ is an artificial coefficient which has an axial distribution, and the electron diffusion coefficient is equivalent to the Bohm diffusion when $\alpha_B = 1$. The distribution of $\alpha_B$ is usually tuned to match the simulated plasma properties with experimental data.

In this study, a 1D simulation is conveniently used to determine the distribution of $\alpha_B$. This 1D simulation is a reduced model of the 2D simulation [17], in which the distributions and gradients in the radial direction are ignored. The distribution of the tuned $\alpha_B$ is shown in **Fig. 2**. $\alpha_B$ is small around the channel exit, and the minimum $\alpha_B$ is set to 0.035 based on a previous study for an SPT-type thruster [18]. The 2D simulation was performed with this $\alpha_B$ distribution.

### 2.3 Electrostatic sheath model

In the discharge channel of a TAL with shifted acceleration zone, the potential of bulk plasma is close to the anode potential, whereas the metallic channel walls are biased at the cathode potential, as shown in Fig. 1. Thus, there exist electrostatic sheaths with potential drops close to the discharge voltage. The nonneutral sheath plasma cannot be handled by the quasineutral hybrid model that is employed in this study. Thus, a sheath model is required to estimate the sheath thickness for the channel walls. Keidar et al. assumed a space-charge-limited (SCL) sheath for a high-power TAL [11]. In this case, the sheath thickness is expressed by the Child–Langmuir law as

$$d^2 = \frac{4\varepsilon_0}{9}\sqrt{\frac{2e}{m_e}}\frac{\phi_s^{3/2}}{i_{i,w}}, \qquad (2)$$

where $d$, $\varepsilon_0$, $\phi_s$, $i_{i,w}$ are the sheath thickness, vacuum permittivity, sheath voltage, and wall ion current density, respectively. However, if the wall ion fluxes are reduced, the $i_{i,w}$ may not reach the SCL current, and Eq. (1) may yield an overestimated sheath thickness.

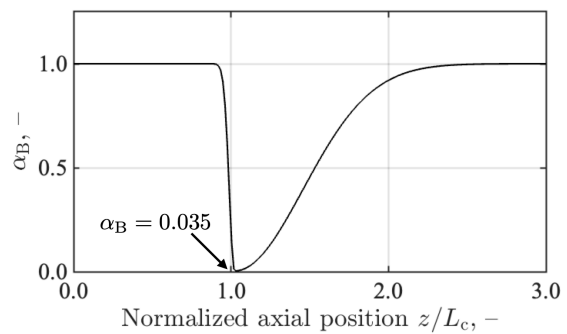

**Fig. 2** Distribution of the coefficient in the Bohm diffusion model determined for RAIJIN66 TAL.

In this study, we considered a one-dimensional model based on collisionless and steady-state sheath with bulk plasma ion, bulk plasma electrons, and secondary electron emission (SEE). The schematics of this sheath model is shown in **Fig. 3**. In the sheath model, the plasma properties at the plasma–sheath edge, i.e. ion flux $\Gamma_{i,s}$, plasma potential $\phi_s$, and electron temperature $T_{e,s}$, are assumed because these quantities are obtained from the hybrid PIC simulation. The Poisson's equation inside the sheath is written as

$$-\frac{\partial^2 \phi}{\partial x^2} = \frac{e}{\varepsilon_0}(n_i - n_e - n_{e,SEE}), \qquad (3)$$

where $\phi$, $n_i$, $n_e$, and $n_{e,SEE}$ are the potential, ion number density, electron number density, and SEE electron density inside the sheath region, respectively. The ions are assumed to enter the sheath with the Bohm velocity, $u_B = \sqrt{eT_e/m_e}$, at the sheath edge, and thus the plasma density at the sheath edge is given as $n_s = \Gamma_{i,s}/u_B$.

The ion flux and energy conservations between the sheath region and sheath edge are expressed as

$$n_i u_i = n_s u_B, \qquad (4)$$

$$\frac{1}{2}m_i u_i^2 + e\phi = \frac{1}{2}m_i u_B^2 + e\phi_s, \qquad (5)$$

where $m_i$ is the ion mass. By using Eqs. (4) and (5), $n_i$ can be expressed as a function of $\phi$ as follows:

$$n_i = n_s\left(1 + 2\frac{\phi_s - \phi}{T_{e,s}}\right)^{-\frac{1}{2}}. \qquad (6)$$

The electrons are assumed to obey the Boltzmann relation inside the sheath, and hence

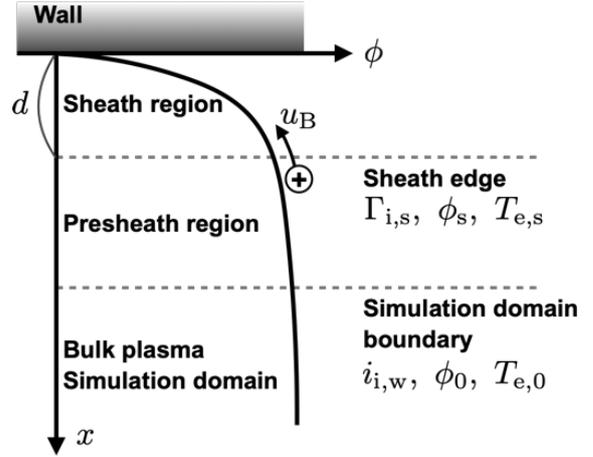

**Fig. 3** Schematics of the 1D sheath model and definitions of plasma parameters at each position.

$$n_e = n_s \exp\left(-\frac{\phi_s - \phi}{T_{e,s}}\right). \qquad (7)$$

SEE electrons are generated when high-energy ions collide with the channel walls. The SEE yield is defined as $\gamma_{SEE} = \Gamma_{e,SEE}/\Gamma_{i,s}$. $\gamma_{SEE}$ is modeled using an empirical expression $\gamma_{SEE} \approx 0.016(E_{ion} - 2E_{work}) = 0.055$ [19], where $E_{ion} = 12.13$ eV is the ionization potential of xenon, and $E_{work} = 4.34$ eV is the work function of stainless steel [20]. $\Gamma_{e,SEE}$ is the flux of SEE electrons, and it is written by a random flux flowing through the surface in one direction as follows:

$$\Gamma_{e,SEE} = \frac{1}{2}n_{e,SEE,w}\sqrt{\frac{8}{\pi}\cdot\frac{eT_{wall}}{m_e}}, \qquad (8)$$

where $T_{wall}$ is the wall temperature in electron volt. The electron flux and energy conservations between the sheath region and wall surface can be defined in a similar manner to Eqs. (4) and (5). Eventually, the distribution of SEE electron density is expressed as

$$n_{e,SEE} = \gamma_{SEE}\Gamma_{i,s}\left(\frac{2e}{\pi m_e}\cdot(T_{wall} + \pi\phi)\right)^{-\frac{1}{2}}. \qquad (9)$$

The $n_i$, $n_e$, and $n_{e,SEE}$ in Eqs. (6), (7), and (9) are inserted into Eq. (3), and the equation can be solved for $\phi$. Because Eq. (3) is a second-order partial differential equation, the equation can be computed by assuming the $\phi$ at the sheath edge and wall as the boundary conditions. Further, by assuming that the gradient of potential at the sheath edge is as small as $(\partial\phi/\partial x)_s \approx 0$, one can calculate the sheath thickness $d$. In this study, $\phi$ and $d$ are calculated iteratively to satisfy the given relations and boundary conditions. The initial value of sheath thickness $d$ is given by

$$d = \left[ \frac{\sqrt{2}}{3} \left( \sqrt{1 + 2\frac{\phi_s}{T_{e,s}}} - 2 \right)^{\frac{3}{4}} + 2\sqrt{2} \left( \sqrt{1 + 2\frac{\phi_s}{T_{e,s}}} - 2 \right)^{\frac{1}{2}} \right] \lambda_D, \quad (10)$$

where $\lambda_D$ is the Debye length at the sheath edge. This is the analytic solution of sheath thickness when one neglects the electron population inside the sheath. When considering the electron population, the sheath thickness is larger than the value given by Eq. (10). Thus, $d$ is gradually increased during the iterative process until the relation $(\partial \phi / \partial x)_s = 0$ is satisfied.

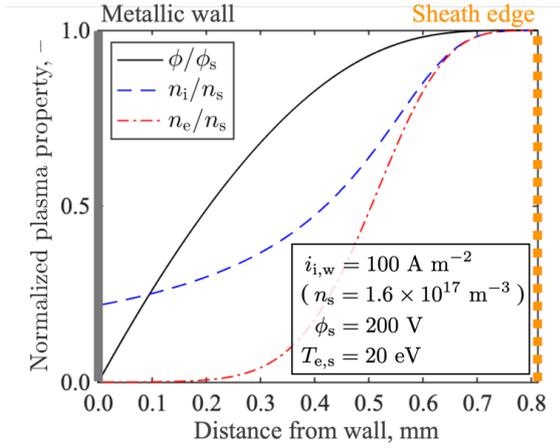

**Fig. 4** Plasma properties inside the channel sheath normalized by the quantities at sheath edge.

One example of the simulated plasma properties inside the sheath is shown in **Fig. 4**. The calculation condition for the sheath edge is also presented in the figure. The plasma properties are normalized by the quantities at the sheath edge. The SEE electron density $n_{e,SEE}$ is less than 0.02% of $n_s$, and thus the effect of SEE electron is negligibly small for the sheath structure. In this example calculation, the sheath thickness was estimated as $d = 0.81$ mm from the 1D sheath model. The use of the Child-Langmuir law in Eq. (1) for the same condition results in an overestimation of sheath thickness as $d = 8.1$ mm. The 1D sheath model is applied to the discharge channel walls.

*2.3 Simulation condition*

The 1 kW level operation of RAIJIN66 TAL [5)] was assumed as the simulation condition. The simulation parameters are presented in **Table 1**. The channel length is defined by the axial distance between the hollow anode tip and channel wall exit. The magnetic field geometry of this thruster is shown in Ref. 5. This thruster is characterized by the magnetic field shifted downstream. Along the channel centerline, the peak of magnetic flux density is located at 2 mm downstream of the channel exit. Simulation domain and boundary conditions are shown in **Fig. 5**. As well as the previous work in Ref. 8, the domain includes the hollow anode and plume regions in the axial direction, and the radial length of the domain is set to hollow anode distance.

**Table 1** Simulation parameters for RAIJIN66 TAL.

| Parameter | Value |
| --- | --- |
| Outer channel diameter | 66 mm |
| Channel length | 2.0 mm |
| Channel width | 12.0 mm |
| Hollow anode distance | 6.0 mm |
| Anode flow rate | 3.4 mg/s |
| Peak magnetic flux density | 30 mT |
| Anode potential | 200 V |
| Cathode-side potential | 50 V |
| Cathode-side electron temperature | 2.6 eV |
| Computational grid | 60×20 |
| Time step for PIC | 10 ns |
| Time step for electron fluid | 0.5 ns |

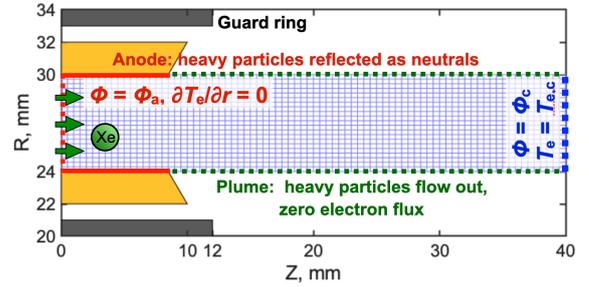

**Fig. 5** Calculation domain with 60×20 grid system and boundary conditions for the axial-radial 2D hybrid PIC simulation.

### 3. Results and Discussion
*3.1 Validation of electron transport model*

**Figure 6** compares the space potential and electron temperature distributions along the channel centerline between the simulation and probe measurement. It is indicated that the main acceleration region is downstream of channel exit. This is the feature of TAL with the shifted acceleration zone. The characteristics of the bulk plasma is close to those of magnetic-layer type SPT, rather than those of conventional TALs. The matching between the simulation and experiment means that the artificial electron transport model is well adjusted.

*3.2 Hybrid PIC simulation results*

**Figure 7** shows the distributions of the simulated plasma density and space potential distributions. Because the present simulation model assumes the quasineutrality, the steep potential drop inside the sheath does not appear in the simulation results. It is observed that the plasma is generated around the channel exit. The peak plasma density is at $z = 13$

Sheath Analysis of Conducting Channel Walls in Anode-Layer Hall Thruster with Reduced Wall Ion Losses

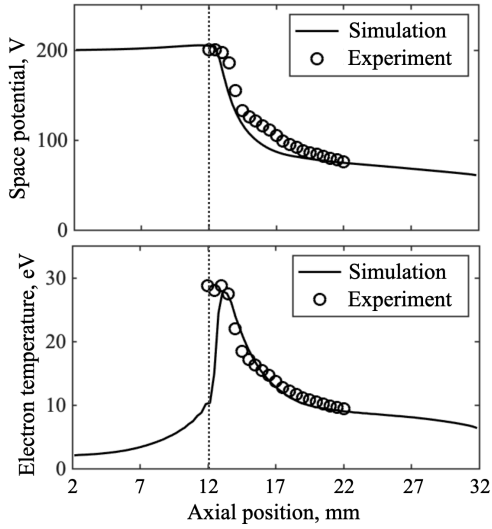

**Fig. 6** Comparisons of space potential and electron temperature distribution between the simulation and probe measurement results.

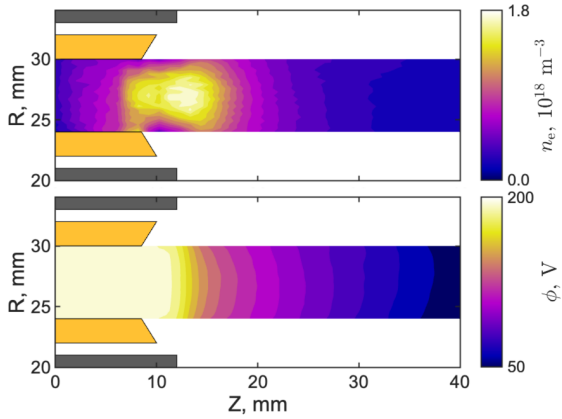

**Fig. 7** Time-averaged distributions of simulated electron number density and space potential.

mm, indicating that most of the generated ions are extracted from the thruster without colliding with the walls. Although the nonneutral sheath is not handled in the present hybrid model, the effect of presheath is observed in the potential distribution. The convex-shaped equipotential line in the discharge channel attracts some ions toward the channel walls. Owing to the presheath, the TAL with shifted acceleration zone admits some ion losses to channel walls.

The wall ion current densities of inner and outer channel walls are presented in **Fig. 8**. The wall ion current density of 100 A/m² was observed in the simulation around the channel exit. In the experiment of RAIJIN66, the guard-ring current was used as an indicator of wall ion losses. The average wall ion current densities that simulated from the present model and that estimated from the measured guard-ring current [22] are compared in **Table 2**. The wall ion current densities are overestimated by 18%. It is inferred that this overestimation is because the wall ion fluxes are evaluated at the top and bottom boundaries ($r = 24$ mm for inner and $r = 30$ mm for outer wall) of the domain, but not at the actual wall positions. Nevertheless, it is assumed that the present simulation reproduces the bulk plasma characteristics in the RAIJIN66, and the sheath thickness is analyzed by the 1D sheath model using the simulated plasma properties.

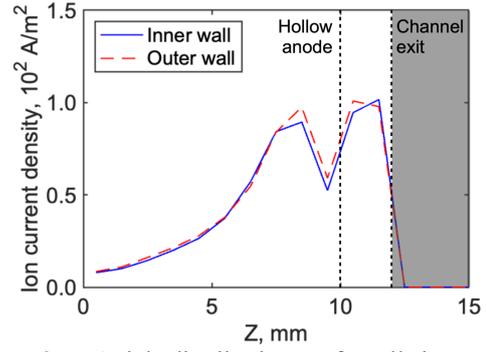

**Fig. 8** Axial distributions of wall ion current density along the inner and outer channel walls.

**Table 2** Wall ion current densities obtained from the numerical simulation and measurement [22].

| Simulation | | Experiment |
|---|---|---|
| Inner ring average | Outer ring average | Inner and outer average |
| 82.9 A/m² | 85.9 A/m² | 70.1 A/m² |

*3.3 Sheath analysis*

Sheath thickness of the guard rings around the channel exit is estimated as $d = 0.64$ mm by using the simulated plasma properties: $i_{i,w} = 100$ A m⁻², $\phi_s = 191$ V, and $T_{e,s} = 9.8$ eV. In this case, one can assume that the sheath region is limited in front of the channel wall. The effects of sheath are insignificant for the bulk plasma, and the main acceleration zone is maintained downstream of the channel exit.

If one considers about the design improvement to reduce the wall ion losses and channel wall erosions, the wall ion current should be reduced. **Figure 9** shows the sheath analysis result in the case of reduced wall ion current density of 5.0 A/m² (1/20 of the simulated ion current density). In this case, the sheath thickness is estimated as $d = 3.0$ mm. This means that the sheath region covers 50% of the channel width, considering the sheaths of the inner and outer channel walls.

One concern about the increased sheath region is that the propellant utilization is decreased. Inside the sheath one cannot expect frequent ionization. Some fraction of the propellant gas supplied from the hollow anode flows through the sheath region without traversing the ionization region. To attain both the reduced wall ion loss and efficient propellant utilization, one must improve the propellant injection method so that most of the propellant gas flow through the main ionization zone around the channel centerline.

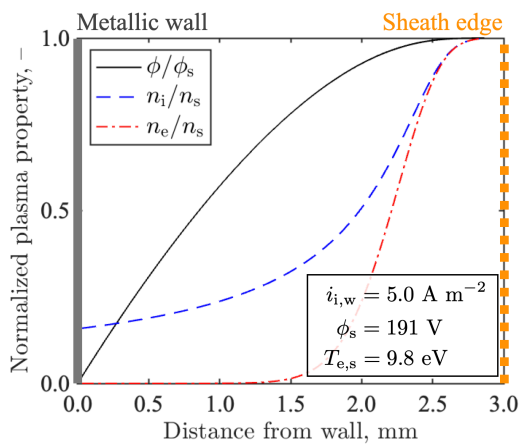

**Fig. 9** Plasma properties inside the channel sheath in the case of reduced wall ion current density.

## 4. Conclusion

Sheath thickness of conducting channel walls in a TAL was analyzed by numerical models. In the TAL with shifted acceleration zone, the formation of thick sheath with large potential gradient is concerned. The bulk discharge plasma flow is modeled by using a quasineutral hybrid PIC model. A one-dimensional sheath model is used to estimate the sheath thickness based on the simulated plasma properties. The major findings are summarized as follows:

(1) The simulated plasma potential and electron temperature agreed with the experimental results. The hybrid model is capable of reproducing bulk discharge plasma characteristics in a TAL with shifted acceleration zone.
(2) The peak wall ion current density reached 100 A/m$^2$, and the estimated sheath thickness for the channel wall was 0.64 mm. The effects of sheath are insignificant for the bulk plasma and the acceleration zone is maintained downstream.
(3) If the wall ion loss is reduced to 5.0 A/m$^2$, the sheath region will cover the 50% of the channel width. Degradation of propellant utilization is concerned in the case of extended sheath region.

**Acknowledgment**

This work was supported by JSPS KAKENHI Grant Number JP20H02346 and Foundation for the Promotion of Ion Engineering.